\documentclass[aps,prd,showpacs,nobibnotes,twocolumn, 
superscriptaddress,nofootinbib]{revtex4}
\usepackage{graphicx}
\usepackage{amsmath,amssymb}
\usepackage{epstopdf}

\begin{document}
\title{Gamma-ray Constraints on Dark Matter Annihilation into Charged
  Particles}

\author{Nicole F. Bell}
\affiliation{School of Physics, The University of Melbourne, Victoria 3010, 
Australia}

\author{Thomas D. Jacques} 
\affiliation{School of Physics, The University of Melbourne, Victoria 3010, 
Australia}

\date{February 5, 2009}

\begin{abstract}
Dark matter annihilation into charged particles is necessarily
accompanied by gamma rays, produced via radiative corrections.
Internal bremsstrahlung from the final state particles can produce
hard gamma rays up to the dark matter mass, with an approximately
model-independent spectrum.
Focusing on annihilation into electrons, we compute robust upper
bounds on the dark matter self annihilation cross section $\langle
\sigma_A v \rangle_{e^+e^-}$ using gamma-ray data from the Milky Way
spanning a wide range of energies, $\sim10^{-3} - 10^4$ GeV.  We also
compute corresponding bounds for the other charged leptons.
We make conservative assumptions about the astrophysical inputs,
and demonstrate how our derived bounds would be strengthened if
stronger assumptions about these inputs are adopted.
The fraction of hard gamma rays near the end point accompanying
annihilation to $e^+e^-$ is only a factor of $\alt 10^2$ lower than
for annihilation directly to monoenergetic gamma rays.  The bound on
$\langle \sigma_A v \rangle_{e^+e^-}$ is thus weaker than that for
$\langle \sigma_A v \rangle_{\gamma\gamma}$ by this same factor.
The upper bounds on the annihilation cross sections to charged leptons are
compared with an upper bound on the {\it total} annihilation cross
section defined by neutrinos.
\end{abstract}



\maketitle

\newpage

\section{Introduction}

A wealth of observational evidence attests to the existence of dark
matter (DM) (see, e.g.,
Refs.~\cite{Kamionkowski_review,Bertone_review,Bergstrom_review} for
reviews).  However, despite the fact that DM accounts for a large
fraction of the total energy density of the Universe, it has evaded
direct detection and its particle properties remain unknown.

If dark matter is a thermal relic, it should have a small, but
non-negligible, self annihilation cross section.  This allows a means
of indirect detection of DM, as annihilations in the Universe today
may generate an observable flux of products.
For thermal relic DM, a total annihilation cross section of $\langle
\sigma_A v \rangle \sim 3\times10^{-26} \textrm{cm}^3 \textrm{
  s}^{-1}$ is needed in order to obtain the observed relic abundance
of $\Omega_{\textrm{DM}} \simeq 0.3$.  (We shall work with the
quantity $\langle \sigma_A v \rangle$, which is the product of the
annihilation cross section and relative velocity, averaged over the
dark matter velocity distribution.  In the Milky Way, $v_{\textrm{rms}} \sim
10^{-3}c$.)  If DM is not a thermal relic, e.g.,
Refs.~\cite{Das:2006ht,Fairbairn:2008fb,KKT}, even larger annihilation
cross sections are possible.  There have also been a number of recent
proposals in which $\langle \sigma_A v \rangle$ is enhanced at low
velocity~\cite{Hisano:2004ds,Feng:2008ya,MarchRussell:2008yu,Cirelli:2008pk,ArkaniHamed:2008qn,Pospelov:2008jd,Nelson:2008hj,Cholis:2008qq,Bai:2008jt,Fox:2008kb}
(i.e., in galactic halos) while still satisfying the thermal relic
constraints. (Note that these scenarios are subject to constraints
arising from annihilations in the first collapsed
structures~\cite{Kamionkowski:2008gj}.)
%

While the total cross section may be related to the relic abundance,
the branching ratios to any particular final states are
model-dependent.  
If we assume the DM is the lightest new particle beyond those in the
standard model (SM), then the branching ratio to SM final states must
be 100\%.  Annihilations must then produce fluxes of detectable
particles emanating from regions of DM concentration, with signals
such as gamma rays, microwaves, neutrinos, and positrons being of
particular interest.
There exist general constraints on the total annihilation
cross-section, based upon unitarity~\cite{Griest:1989wd, Hui:2001wy}
and the requirement that annihilations not significantly alter halo
density profiles~\cite{KKT}.  A strong bound is placed on the total
annihilation cross section by assuming that the branching ratio to
neutrinos, the least detectable final state, is 100\%~\cite{BBM,YHBA}.

We focus here on the process $\chi\chi\rightarrow e^+ e^-$ in which DM
annihilates to an electron-positron pair, though we shall also report
results for the other charged leptons.
Although the branching ratio to this particular final state is
model-dependent, it is a significant channel in a wide range of
models.  For example, while annihilation to fermions is helicity
suppressed in supersymmetric models, Kaluza-Klein DM features large
(unsuppressed) annihilation rates to leptons \cite{Servant:2002aq,Cheng:2002ej,Hooper:2004xn}, as does the Dirac DM model of Ref. \cite{Harnik:2008uu}.  
Numerous authors have recently proposed models in which annihilation
to charged leptons is significantly
enhanced~\cite{Cirelli:2008pk,ArkaniHamed:2008qn,Pospelov:2008jd,Nelson:2008hj,Cholis:2008qq,Bai:2008jt,Fox:2008kb},
making upper limits on the cross section to these annihilation
products particularly interesting.
%
%
In addition, various other SM final states, such as $W^+W^-$ and $ZZ$,
produce $l^+l^-$ via their decays and hence a flux of charged leptons
is of generic interest in a large variety of DM models.

Several techniques may be used to constrain the production of $e^+e^-$
within galactic halos, all of which rely on the fact the charged
particles inevitably produce photons.  Signals considered include
gamma rays, x-rays, microwaves and radio waves.
Photons are produced by the various energy loss processes that charged
particles undergo in a galactic halo, examples of which include
synchrotron radiation due to the propagation of $e^\pm$ in galactic
magnetic fields, and inverse Compton scattering of electrons from
interstellar radiation fields,
e.g.,~\cite{Eloss1,Eloss2,Eloss3,Aloisio:2004hy,Hooper:2007kb,Hooper:2008zg,Borriello:2008gy,Zhang:2008rs,Grajek:2008jb,Jeltema:2008ax}.
The drawback of these techniques is a significant dependence on
astrophysical inputs, some of which are poorly known.  Uncertainties
in magnetic field strengths, radiation backgrounds, and electron
diffusion scales all enter the calculations in an involved fashion.

Charged particles also produce photons via electromagnetic radiative
corrections~\cite{BBB,Birkedal:2005ep,BergstromRC,Bergstrom:2005ss,Bergstrom1,Bergstrom2,Bergstrom3,Bergstrom4,Schleicher:2008gm}.
The lowest order dark matter annihilation process $\chi\chi
\rightarrow e^+e^-$ is necessarily accompanied by the radiative
correction $\chi\chi \rightarrow e^+e^- \gamma$.  This is an {\it
  internal bremsstrahlung} (IB) process, meaning that the photon
arises at the Feynman diagram level and is not due to interaction of
charged particles in a medium.  Importantly, for a given annihilation
cross section, $\langle \sigma_A v \rangle_{e^+e^-}$, the accompanying
flux of IB photons can be determined without knowledge of the new
underlying particle physics which mediates the DM annihilation.
Moreover, IB suffers none the drawbacks of the competing inverse
Compton and synchrotron techniques outlined above.
While inverse Compton and synchrotron fluxes are dependent on
conditions of the astrophysical environment, the IB flux is always
present and its normalization and spectrum are predetermined.

In this paper we use IB emission to derive robust upper limits on the
dark matter annihilation cross section to electron-positron pairs
$\langle \sigma_A v \rangle_{e^+e^-}$ over a wide DM mass range spanning
$\sim10^{-3}-10^{4}$ GeV.  We calculate DM annihilation fluxes produced in
the galactic halo, and compare with the gamma-ray backgrounds reported
by COMPTEL, EGRET and H.E.S.S.  We also look at data for the M31
(Andromeda) galaxy, to fill a gap between the energy ranges covered by
EGRET and H.E.S.S.
We explicitly demonstrate how our limits vary according to the assumed
DM halo profile (our one source of uncertainty) and also compare our
limit on the annihilation cross section to $e^+e^-$ with corresponding
bounds on the $\gamma\gamma$ and $\bar\nu\nu$ final states.

\section{Internal Bremsstrahlung}

\begin{figure}
\includegraphics[width=3.0in, clip=true]{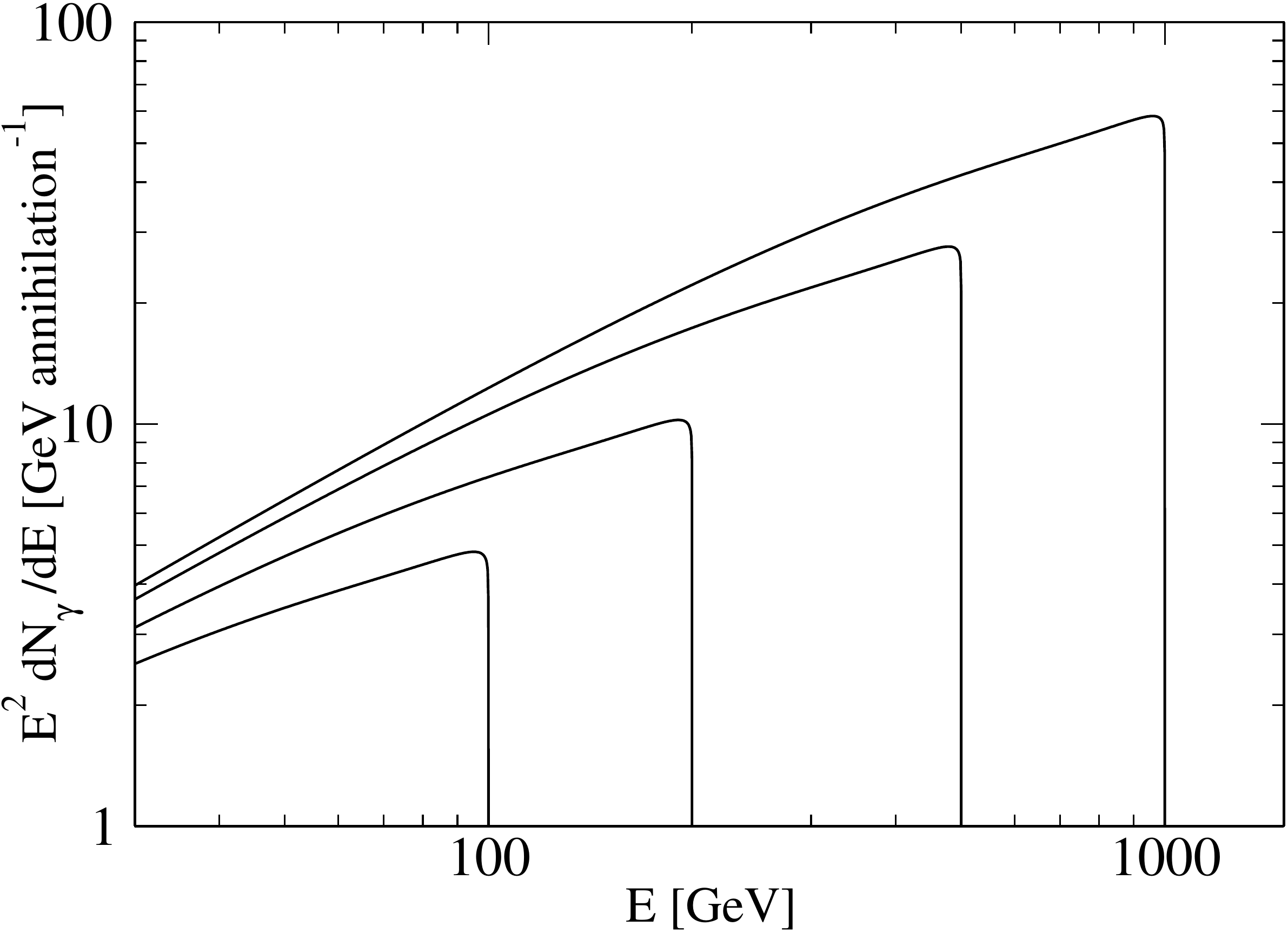}
\caption{ Internal bremsstrahlung gamma-ray spectra per
  $\chi\chi\rightarrow e^+e^-$ annihilation, for $m_\chi=$ 100,
  200, 500, and 1000 GeV.}
\label{spectra}
\end{figure}

If DM annihilates to produce charged particles, the lowest order
processes will always be subject to electromagnetic radiative
corrections, resulting in the production of real photons.  In
particular, the annihilation $\chi\chi\rightarrow e^+e^-$ will be
accompanied by the internal bremsstrahlung process
$\chi\chi\rightarrow e^+e^-\gamma$.  A photon may be emitted from
either the final state $e^+$ or $e^-$, with a cross-section
proportional to $\alpha\simeq 1/137$.  See
Refs.~\cite{PS,Birkedal:2005ep,BBB,BergstromRC} for a detailed discussion.  To a
good approximation, the differential cross section for $\chi\chi
\rightarrow e^+e^-\gamma$ is
\begin{equation}
\frac{d\sigma_{\rm IB}}{dE} = \sigma_{\rm tot} \times 
\frac{\alpha}{E \pi} 
\bigg[ \ln\bigg(\frac{s'}{m_e^2} \bigg) -1 \bigg]\bigg[1+\bigg( \frac{s'}{s}\bigg)^2\bigg],
\label{sigbrem}
\end{equation}
where $E$ is the photon energy, $s=4m_\chi^2$,
$s'=4m_\chi(m_\chi-E)$, and $\sigma_{\rm tot}$ is the
tree-level cross section for $\chi\chi \rightarrow e^+e^-$.  Note that
$\sigma_{\rm tot}$ factors from the IB cross-section.  This
important feature implies that the IB spectrum is independent of the
unknown physics which mediates the lowest order annihilation process.
The photon spectrum per $\chi\chi \rightarrow e^+e^-$ annihilation is
therefore given by
\begin{equation}
\frac{dN_\gamma}{dE} = 
\frac{1}{\sigma_{\rm tot}}\frac{d\sigma_{\rm IB}}{dE_\gamma}.
\label{Ngamma}
\end{equation}
This spectrum is shown in Fig.~\ref{spectra} for various choices of the
DM mass, where a sharp edge in the spectrum at $E =m_\chi$ is
evident.

Note that we consider only radiation from the final state particles,
and not from any internal propagators.  In some supersymmetric
scenarios in which $s$-channel annihilation to fermions is helicity
suppressed, bremsstrahlung from internal propagators can be
particularly important as it can circumvent this
suppression~\cite{Bergstrom1,Bergstrom2,Bergstrom3,Bergstrom4}.  We do
not consider these model-dependent processes.  Note, however, that the
presence of such emission would only increase the gamma-ray flux we
calculate, and hence strengthen the cross section limits derived.

\section{Annihilation in Dark Matter Halos}

The rate at which dark matter annihilates is proportional to the
square of the dark matter number density,
$n_\chi=\rho/m_\chi$.  However, there are considerable uncertainties
in the dark matter density profile of the galactic halo, particularly
in the central region where the density, and hence the annihilation
rate, is largest.  To deal with these uncertainties, we make
conservative assumptions about the DM density profile, and show how
our results vary if less conservative assumptions were used.

A standard parameterization of a dark matter halo density profile is 
\begin{equation}
\rho(r) =
\frac{\rho_0}{\left(r/r_s\right)^{\gamma}
\left[1+\left(r/r_s\right)^\alpha\right]^{(\beta - \gamma)/\alpha}}\,.
\end{equation}
The Kravtsov \cite{Kravtsov:1997dp}, Navarro-Frenk-White (NFW)
\cite{NFW} and Moore \cite{Moore:1999gc} profiles are defined by the
values $(\alpha,\beta,\gamma)$ shown in Table~\ref{profiles}.
%
%
The Milky Way values for $r_s$, the scale radius, and $\rho(R_{\rm
  sc})$, the DM matter density at the solar circle $R_{\rm sc}=8.5$ kpc, are also given in Table~\ref{profiles}.  The
normalization $\rho_0$ is then fixed by $\rho(R_{\rm sc})$ and
$r_s$.
For large radii $r\agt r_s$, all three profiles scale with radius as
$1/r^3$ and are normalized such that they coincide closely.  However,
the profiles diverge for small radii, scaling as $1/r^{0.4}$, $1/r$
and $1/r^{1.5}$ for the Kravtsov, NFW and Moore profiles respectively.
The steep Moore profile thus features a greatly enhanced density near
the Galactic center, compared to the relatively flat Kravtsov profile;
the NFW profile falls between the two.


\begin{table}
\caption{Values of the parameters $(\alpha, \beta,\gamma)$ which define
  the Kravtsov, NFW and Moore halo profiles.  The values of the scale
  radius $r_s$ in [kpc], and halo density normalization at the solar
  circle $\rho(R_{\rm sc})$ in [GeV cm$^{-3}$], are specific to the
  Milky Way.  }
\label{profiles}
\begin{center}
  \begin{tabular}{cccccc}
    \hline\hline
    & $\,\, \alpha \,\, $  & $\,\, \beta \,\,$ & $\,\, \gamma \,\,$ 
    & $\,\, r_s \,\,$ & $\,\, \rho(R_\textrm{sc})$   \\
    \hline
    Kravtsov & 2 & 3 & 0.4 & 10 & 0.37  \\
    NFW      & 1 & 3 & 1   & 20 & 0.3   \\
    Moore & 1.5 & 3 & 1.5 & 28 & 0.27 \\
    \hline\hline
\end{tabular}
\end{center}
\end{table}

Uncertainties in the density of the DM halo translate into
uncertainties in the DM annihilation rate.  A detailed discussion of
the dependence of DM annihilation signals on the choice of density
profile is given in Ref.~\cite{YHBA}.  While the uncertainties are
mild for large angular regions of the galaxy, they scale several
orders of magnitude for small angular scales close to the Galactic
center.  In order to place conservative upper limits on the DM
annihilation cross section, we focus on the profile with the smallest
dark matter density, namely the Kravtsov profile.  For the NFW and
Moore profiles, smaller values of $\langle \sigma_A v\rangle$ are
needed to reproduce the same flux, and hence lead to stronger (less
conservative) limits.

We now calculate the gamma-ray flux for annihilations in the galactic
halo.  For an observation direction at angle $\psi$ with respect to
the Galactic center, the integral of the square of the dark matter
density along the line of sight is given by
\begin{equation}
{\cal J}(\psi) = {\rm J_0}
\int^{\ell_{max}}_0 \rho^2\left(\sqrt{R_{\textrm{sc}}^2 -
2\ell R_{\textrm{sc}}\cos{\psi} +\ell^2}\right)d\ell \,,
\label{los}
\end{equation}
where J$_0=1/[8.5\,{\rm kpc} \times(0.3 \,{\rm GeV \, cm}^{-3})^2]$ is
an arbitrary normalization constant used to make $\cal{J}(\psi)$
dimensionless, and which cancels in our final expression for the
gamma-ray flux. We then define the average of $\cal{J}(\psi)$ over an
observation region of solid angle $\Delta\Omega$ as
\begin{equation}
{\cal J}_{\Delta \Omega} = \frac{2\pi }{\Delta \Omega} \int_0^{\psi}
{\cal J}(\psi )\sin{\psi }d\psi.\,
\end{equation}
Values of ${\cal J}(\psi)$ and ${\cal J}_{\Delta \Omega}$ are given
$\psi$ in Fig.~2 of Ref.~\cite{YHBA}.
With these definitions, the gamma-ray flux per
steradian due to DM annihilation in an observation region of angular
size $\Delta\Omega$ is
\begin{eqnarray}
\frac{d\Phi_{\gamma}}{dE} &=& \frac{\langle \sigma_A v \rangle}{2} \frac{\mathcal{J}_{\Delta\Omega}}{4 \pi m_\chi ^2{\rm J}_0}\frac{dN_\gamma}{dE}\label{flux1},
\end{eqnarray}
where $dN_\gamma/dE$ is the gamma-ray spectrum per annihilation.  
For the IB emission associated with annihilation to $e^\pm$, we must
replace $\langle \sigma_A v \rangle$ with $\langle \sigma_A v
\rangle_{e^+e^-}=\langle \sigma_A v \rangle\times {\rm Br}(e^+e^-)$,
while $dN_\gamma/dE$ is given by Eq.~\ref{Ngamma}.

\section{Analysis of Annihilation flux}

\subsection{Analysis technique}

Our analysis technique is similar to that followed in Mack \emph{et
  al.}~\cite{MJBBY}.  We use galactic gamma-ray data from
COMPTEL~\cite{COMPTELweb}, EGRET~\cite{EGRETweb} and
H.E.S.S.~\cite{HESSweb}, together spanning the broad energy range
$10^{-3}-10^4$ GeV.  As there is a small gap between the energy ranges
covered by EGRET and H.E.S.S., we use the observations of the M31
(Andromeda) galaxy made by CELESTE~\cite{CELESTEweb} to calculate
constraints for this energy interval.

The galactic gamma-ray background measurements reported by COMPTEL,
EGRET and H.E.S.S. are given in approximately log-spaced energy intervals,
with energy bins of size ranging from
$\Delta \log E \sim 0.2 - 0.6$.
We calculate the IB gamma-ray flux for the observation regions viewed
by these experiments, using the methods outlined above, and compare
with the observational data.  Upper limits on $\langle \sigma_A v
\rangle_{e^+e^-}$ are determined by requiring that the IB flux due to DM
annihilation be lower than 100\% of the observed gamma-ray background
flux in each of the experimental energy bins.
Given that a large fraction of the observed gamma-ray background is
likely to be astrophysical in origin, and not due to DM annihilation,
taking the total background flux to be an upper limit on the DM
annihilation signal is an extremely conservative approach.

\begin{figure}[t]                                                             
\includegraphics[width=3.5in, clip=true]{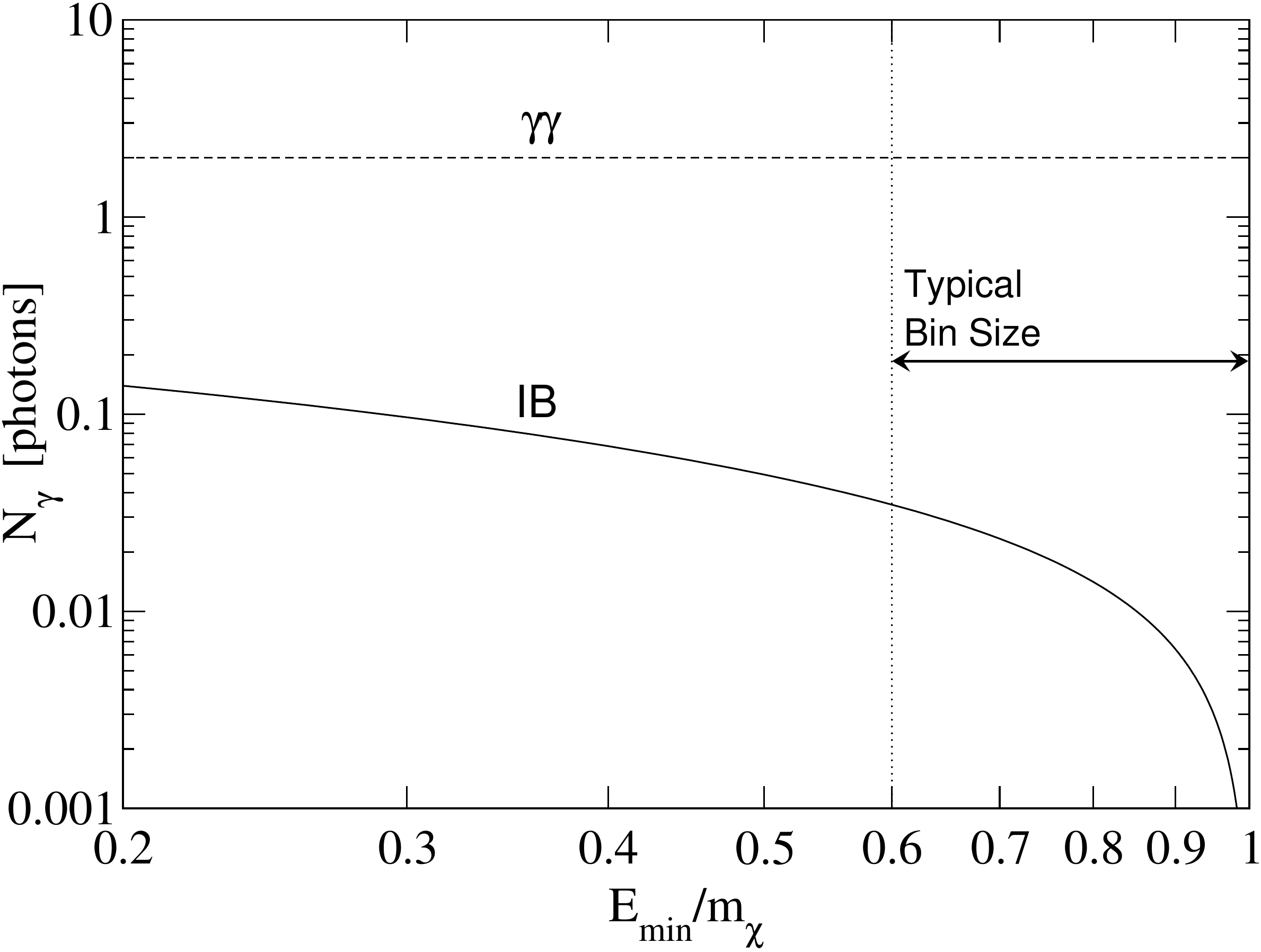}                      
\caption{Number of gamma rays per DM annihilation ($\int_{E_{\rm
      min}}^{m_\chi}dE\,\, dN_\gamma/dE$) as a function of the lower
  limit of integration, for IB emission from $\chi\chi\rightarrow
  e^+e^-$ (solid line).  A typical bin size used in the analysis is
  shown.  The DM mass used is 1000 GeV; variation with $m_\chi$ is
  very small.  Shown for comparison is the number of photons per
  annihilation for the process $\chi\chi\rightarrow\gamma\gamma$
  (dashed line) in which the photons are always at the endpoint.}
\label{NvsBinSize}                                                            
\end{figure}                                                                  

For each energy bin, we take the DM mass to be equal to the upper
energy limit of the bin, and integrate the IB flux over the width of
the bin.  Figure~\ref{NvsBinSize} shows the quantity
\begin{equation}                                                              
\int_{E_{\rm min}}^{m_\chi} \frac{dN_\gamma}{dE} dE 
= \frac{1}{\sigma_{\rm tot}} \int_{E_{\rm            
min}}^{m_\chi} \frac{d\sigma_{\rm                   
IB}}{dE}dE,                                                                   
\end{equation}             
which is the number of photons per annihilation as a function of bin
size, for IB emission from $e^\pm$, and $m_\chi=1000$ GeV.  [Since
$m_\chi$ enters Eq.~\ref{sigbrem} via a logarithm, variation with
$m_\chi$ is only very mild.]  This indicates how the size of the energy
bins affects our results.  For comparison, the number of photons per
annihilation for the process $\chi\chi \rightarrow \gamma\gamma$ (for
which $dN_\gamma/dE=2\delta(m_\chi-E)$) is also shown.
Note that the IB cross section is proportional to $1/E$, so for
sufficiently low photon energy the IB probability becomes large and
one must account for multiple photon emission.  However, we are not
working in this regime, and in fact obtain our limits using the flux
near the endpoint of the spectrum.  Despite the fact that the IB flux
is small, we have enough hard gamma rays near the endpoint to result
in strong bounds.
For typical parameters, the IB flux per annihilation to $e^\pm$ is
smaller than the photon flux per annihilation for $\chi\chi
\rightarrow \gamma\gamma$ by a factor of $10^2$.  This is expected,
given that the IB cross section is suppressed by a factor of $\alpha$
with respect to the tree-level DM annihilation process.

\subsection{Observational data}

COMPTEL and EGRET are telescopes aboard the Compton Gamma-Ray
Observatory. Refs.~\cite{Strong:1998ck, Strong:2005zx} present
flux data from these telescopes over an energy range of 1 MeV to 100
GeV between them, with a observation region of $-30^\circ < l <
30^\circ$ in Galactic longitude, and $-5^\circ < b < 5^\circ$ in
Galactic latitude.  We calculate $\mathcal{J}_{\Delta\Omega}$ as if
the DM annihilation signal were from a circular region of
$\psi=30^\circ$.  This is a conservative approach, as the average
annihilation flux per steradian for the circular region is lower than
for the rectangular region which was actually observed, and yields
values of $\mathcal{J}_{\Delta\Omega}=(13,28,100)$ for the Kravtsov,
NFW and Moore profiles respectively.

The H.E.S.S. observations reported in Ref. \cite{Aharonian:2006au}
cover the relatively small angular region $-0.8^\circ < l <
0.8^\circ$, $-0.3^\circ<b<0.3^\circ$, in Galactic coordinates, over an
energy range of 300 to 15 000 GeV.  The data include a background
subtraction which we must take into account when calculating
$\mathcal{J}_{\Delta\Omega}$; see Mack \emph{et al.}~\cite{MJBBY} for
details.  This leads to $\mathcal{J}_{\Delta\Omega}=(3,850,50\,000)$
for the Kravtsov, NFW and Moore profiles respectively.

CELESTE viewed the M31 galaxy with an observation region of angular
radius 0.29$^\circ$.  No signal was seen, and a 2-$\sigma$ upper limit
on the flux of gamma rays from M31 between 50 and 700 GeV of around
$10^{-10}$ photons cm$^{-2}$ s$^{-1}$ was
reported~\cite{Lavalle:2006rs} .
We compare this flux with the IB signal calculated for an energy bin
of size $10^{-0.4} m_\chi - m_\chi$. This is extremely
conservative, as we are constraining the cross section by requiring
that the annihilation flux in a small bin be less than or equal to the
observed flux in a much larger bin.
The DM density profile of M31 is less well constrained than that of
the Milky Way.  As in Ref.~\cite{MJBBY}, we model the M31 halo using
the Kravtsov profile for the Milky Way, which yields
$\mathcal{J}_{\Delta\Omega}\times\Delta\Omega\simeq2\times 10^{-3}$.

\section{Discussion}

  \begin{figure}
\includegraphics[width=3.5in, clip=true]{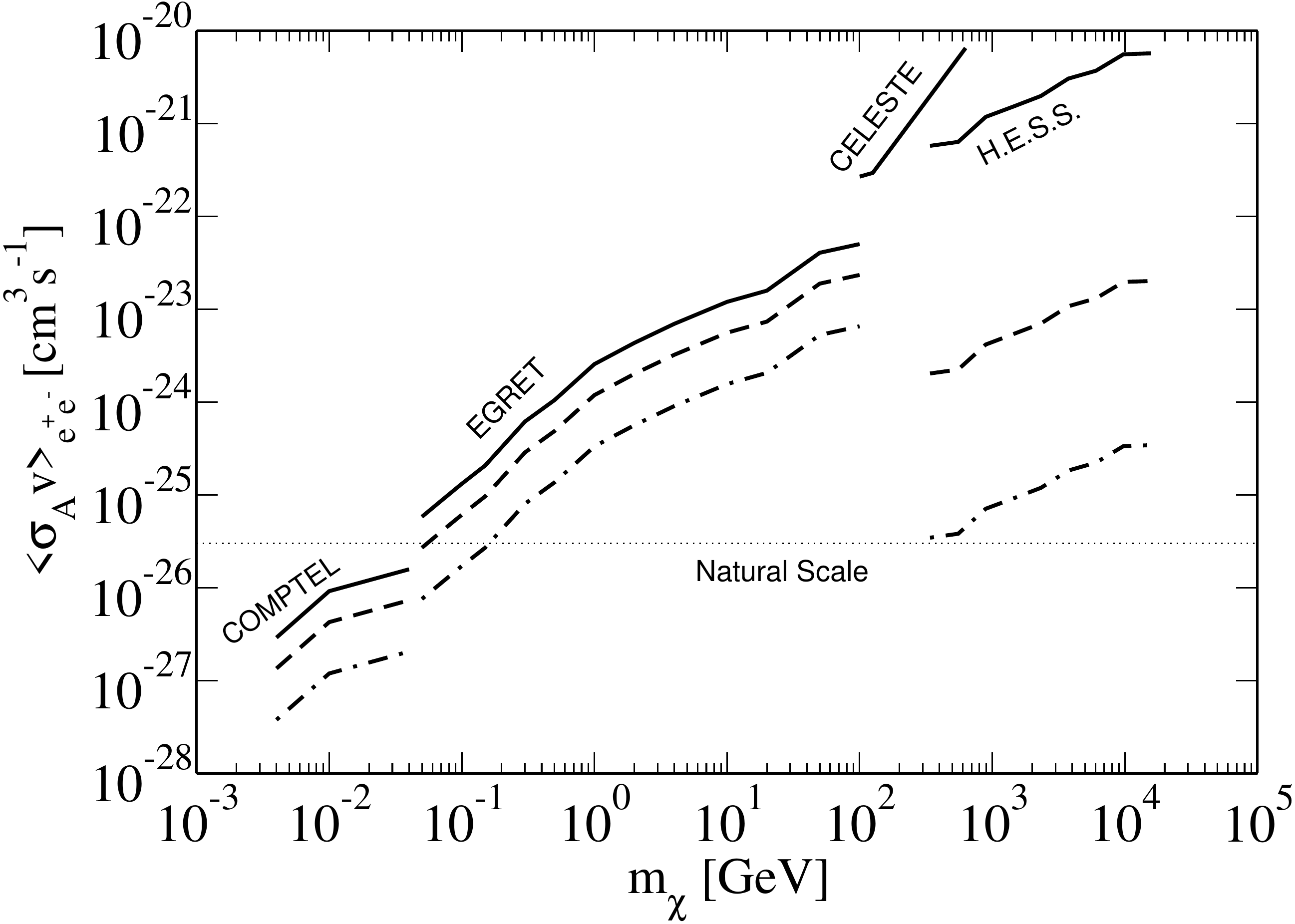}
\caption{Upper limit on $\langle \sigma v \rangle_{e^+e^-}$ as a
  function of DM mass for the Kravtsov (solid line), NFW (dashed line) and
  Moore (dotted-dashed line) profiles.}
\label{results1}
\end{figure}

In Fig.~\ref{results1} we show the upper limits on $\langle \sigma_A v
\rangle_{e^+e^-}$ as a function of DM mass, using the observational
data described above.  We give the Galactic genter results for the
conservative Kravtsov profile, the more commonly adopted NFW profile,
and the steep Moore profile. For the CELESTE observation of M31,
differences between these profiles are expected to have a modest
effect on the results, as a large portion of the galaxy is within the
field of view; in Fig.~\ref{results1} we show the CELESTE constraint
using only the Kravtsov profile.
As previously discussed, while the Kravtsov, NFW and Moore profiles
diverge towards the center of the Galaxy, they are similar at large
radii. As the EGRET and COMPTEL observations encompass relatively
large angular scales, the density profile changes have a modest
effect.  On the other hand, the H.E.S.S. constraints correspond to a much
smaller angular region toward the Galactic center, and vary by orders
of magnitude depending on the profile adopted.
(See Ref.~\cite{YHBA} for a full discussion of the differences between
the profiles for different angular regions.)
%
%
%
To be conservative, we do not consider the possibility that DM
annihilation rates are enhanced due to substructure in the halo, e.g.,
Refs.~\cite{Diemand:2006ik,Strigari:2007at,Bi:2005im,Hooper:2007be},
or mini-spikes around intermediate-mass black
holes~\cite{Bertone:2005xz, Horiuchi:2006de}; such enhanced
annihilation signals would result in stronger upper bounds on the
cross section.

  \begin{figure}
\includegraphics[width=3.5in, clip=true]{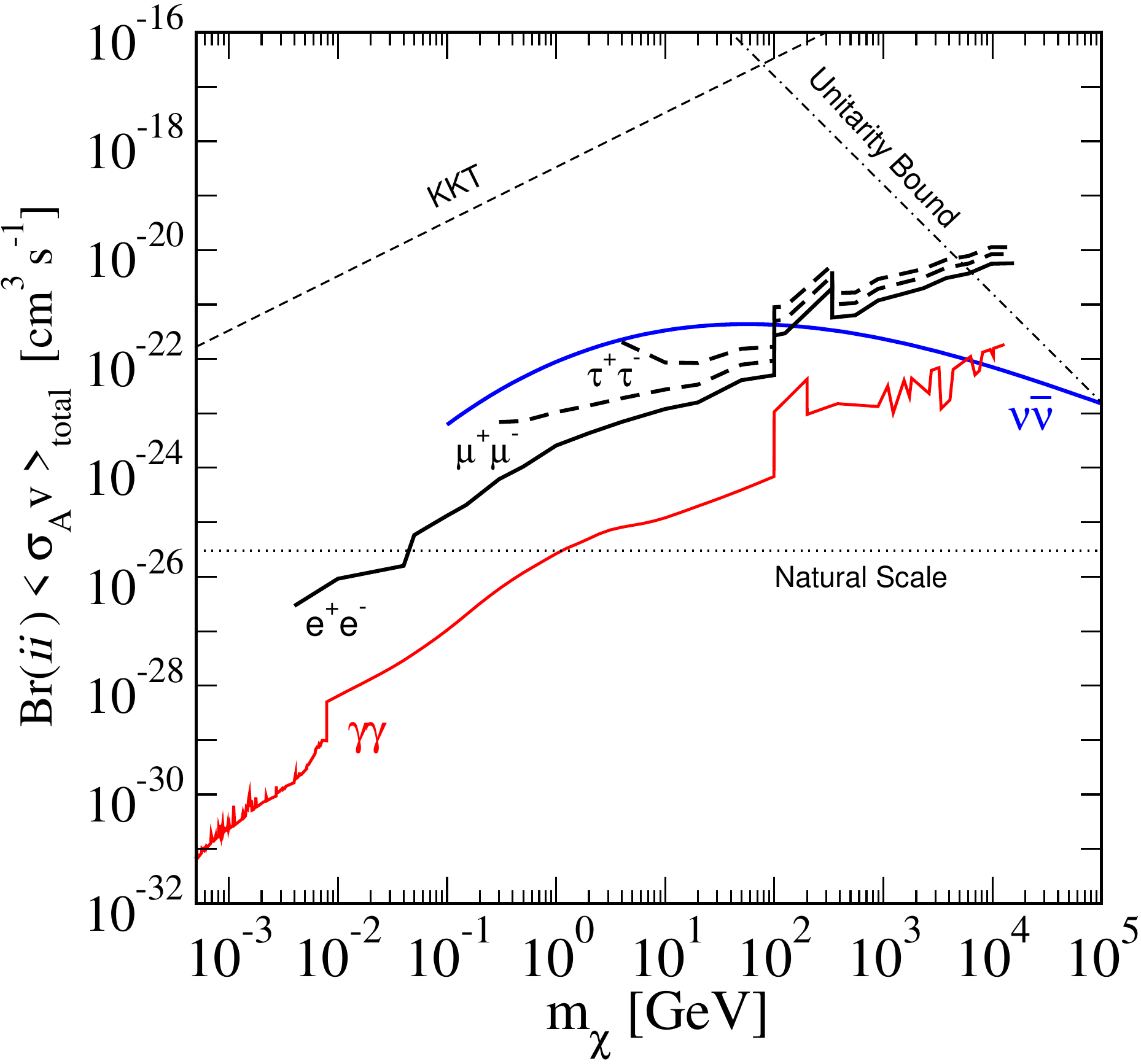}
\caption{Upper limits on the {\it partial} cross sections $Br(ii)\times\langle \sigma v
\rangle_{total}$ for various final states $ii=e^+e^-$ (solid black line;
 labeled), $\mu^+\mu^-$ (thick dashed line; labeled), $\tau^+\tau^-$ (thick
 dashed line; labeled), $\gamma\gamma$ (red line; labeled), and $\bar\nu\nu$
 (blue line; labeled), using the conservative Kravtsov profile.
Each of these partial cross-section limits is independent, with no relationship assumed between the branching ratios to particular final states.
Also shown are the KKT (thin dashed line) and unitarity (thin dotted-dashed line)
 limits on the {\it total} cross section described in the text, and the
 cross section for thermal relic DM (natural scale).  The
 $\gamma\gamma$ and $\bar\nu \nu$ limits are taken from
 Ref.~\cite{MJBBY} and Ref.~\cite{YHBA}, respectively.}
\label{results2}
\end{figure}

We can estimate the way that the cross-section bounds scale with the DM
mass.
The spectrum of the galactic gamma-ray background falls off with
energy as $d\Phi/dE \sim E^{-\alpha}$ where, e.g., $\alpha$ is
slightly larger than 2 in the EGRET and HESS energy ranges.  The IB
signal scales approximately as $d\Phi/dE \sim \langle \sigma_A
v\rangle m_\chi^{-2} E^{-1}$ [where accounting the full energy
dependence in Eq.~\ref{sigbrem} has only a small effect on this scaling].
Given this scaling with $E$, it is clear that the strongest
constraints arise from the endpoint of the spectrum.  In addition,
small bin size is optimal for obtaining strong constraints.
If we integrate the flux over an energy bin of width $x m_\chi$ to
$m_\chi$, the IB flux within the energy bin is proportional to $\langle
\sigma_A v\rangle m_\chi^{-2}$ (ignoring logarithmic corrections) while the
background flux is proportional to $m_\chi^{-\alpha+1}$ (for fixed $x$).  The
cross-section limits then scale with $m_\chi$ as $\langle \sigma_A v
\rangle \sim m_\chi^{3-\alpha}$ and thus rise as $m_\chi$ if $\alpha
\sim 2$.

In Fig.~\ref{results2} we show the upper bounds on the annihilation
cross sections into $e^+e^-$, $\mu^+\mu^-$, and $\tau^+\tau^-$, based
upon the IB emission from each final state (all use the conservative
Kravtsov profile).
As the mass of the charged lepton increases, the rate of IB emission
decreases and thus the upper bounds on the cross sections become
weaker.  However, as the IB spectrum depends only logarithmically on
the charged lepton mass, this effect is mild, particularly for large
DM mass.
We have used Eq.~\ref{sigbrem} to calculate the IB flux over the
entire mass range, and present limits which range from high $m_\chi$
down to just above threshold.  We expect modifications to
Eq.~\ref{sigbrem} in the limit that the charged leptons are
nonrelativistic, but this will only affect a small mass range close
to threshold.
%
%
%
In the case of annihilation to $\tau^\pm$, we have not considered the
gamma rays that arise from hadronic decay modes of the $\tau$ leptons
(see, e.g., Ref.~\cite{BergstromRC}) which form a broad spectrum
centered on the pion mass.

Using Fig.~\ref{results2}, we may compare the limit on the
annihilation cross section into $e^+e^-$ with that for $\gamma\gamma$,
taken from Ref.~\cite{MJBBY}.
As anticipated, the bound on $\langle \sigma v \rangle_{e^+e^-}$ from
IB is weaker than the bound on $\langle \sigma v
\rangle_{\gamma\gamma}$ by a factor of $\sim 10^{-2} \sim \alpha$.
This difference can be understood by comparing the number of hard
gammas near the endpoint, $E_\gamma=m_\chi$.  For annihilation to
$\gamma\gamma$, there are always two monoenergetic gamma rays at the
endpoint.  The IB spectrum, integrated over a typical bin width (see
Fig.~\ref{NvsBinSize}) results in a flux that is smaller than this by
only a factor of less than 100.  That IB provides a limit this close
to the ideal $\gamma\gamma$ channel illustrates the importance of the
IB technique.

We stress that our IB limits apply to the {\it partial} DM
annihilation cross sections, $\langle \sigma_A v \rangle_{l^+l^-}$,
rather than the {\it total} annihilation cross section, $\langle
\sigma_A v \rangle$.  The two are related via $\langle \sigma_A v
\rangle_{l^+l^-} = Br(l^+l^-) \langle \sigma_A v \rangle$.
While the branching ratios are entirely dependent on the choice of DM
model, there are many scenarios which feature large branching ratios for
direct annihilation to leptons, such as Kaluza-Klein DM, or the recent
models of Refs.~\cite{Cirelli:2008pk,ArkaniHamed:2008qn,Pospelov:2008jd,
  Nelson:2008hj,Cholis:2008qq,Bai:2008jt,Fox:2008kb}.
Our constraints can be readily applied to any particular
model, simply by dividing the $\langle \sigma_A v \rangle_{l^+l^-}$
bounds by the relevant branching ratios.

Also shown in Fig.~\ref{results2} are a number of upper bounds on the
{\it total} annihilation cross-section, based upon
unitarity~\cite{Griest:1989wd, Hui:2001wy} and the requirement that DM
annihilation not significantly alter halo density profiles
[Kaplinghat-Knox-Turner (KKT)~\cite{KKT}].
Note that the limit on annihilation to $\bar\nu\nu$ also defines a
strong bound on the {\it total} DM annihilation cross
section~\cite{BBM}.  If we assume annihilation to only standard model
final states, a conservative bound on the total cross section is
obtained by assuming the branching ratio to neutrinos (the least
detectable final state) is 100\%.  Any other assumption would lead to
appreciable fluxes of gamma rays and hence be more strongly
constrained.
Dark matter annihilation into neutrinos was examined in
Refs.~\cite{BBM} and~\cite{YHBA}.  The upper bound on $\langle
\sigma_A v \rangle_{\bar\nu \nu}$ shown in Fig.~\ref{results2} (taken
from~\cite{YHBA}) shows that the neutrino constraints are quite
strong, particularly for large DM mass.
(Reference~\cite{Sergio} extended the DM annihilation limits to lower
masses, while Ref.~\cite{Yin:2008mv} included substructure
enhancement.  Reference~\cite{Sergio_decay} derived analogous limits on the
DM decay rate.)
In fact, due to electroweak bremsstrahlung (radiation of $W$ and $Z$
bosons, rather than photons) neutrinos are also inevitably accompanied
by photons~~\cite{Kachelriess:2007aj,BDJW,Dent:2008qy}.  However, this
results in bounds that are comparable to, or weaker than, those
obtained directly from neutrinos.

If DM annihilates to $e^\pm$, photons will be produced not only by IB,
but also by energy loss processes including inverse Compton scattering
and synchrotron radiation.  In particular, radio wavelength signals
produced via synchrotron emission have been the focus of much recent
attention,
e.g.,~\cite{Hooper:2007kb,Hooper:2008zg,Borriello:2008gy,Zhang:2008rs,Grajek:2008jb}.
However, the intensity of synchrotron radiation depends on a number of
uncertain astrophysical parameters, such as magnetic field strength,
radiation field intensities, and electron diffusion scales.  By
contrast, IB is free of these astrophysical uncertainties, and has a
fixed spectrum and normalization.
Another key difference is the energy of the photons.  Synchrotron radiation
produces generally low energy photons, while IB provides some hard
gamma rays near the endpoint.  Since the background flux falls off
with energy, these hard gamma rays are extremely useful.
The sharp edge in the IB spectrum at $E=m_\chi$ can be used to
diagnose the DM mass; this is not possible with synchrotron radiation.

Nonetheless, it useful to take the synchrotron-based cross-section
bounds as a reference point to compare with our IB-based bounds.  Our
conservative IB bound on $\langle \sigma_A v \rangle_{e^+e^-}$ is
comparable to conservative bounds on $\langle \sigma_A v
\rangle_{W^+W^-}$ obtained from synchrotron radiation.  For example,
Ref.~\cite{Grajek:2008jb} obtains $\langle \sigma_A v \rangle_{W^+W^-}
\alt 4 \times 10^{-24} \textrm{cm}^3 \text{s}^{-1}$
$(4 \times 10^{-23} \textrm{cm}^3 \text{s}^{-1})$ at 100
GeV (1 TeV) assuming an NFW profile and conservative magnetic field
choices (lower panel of Fig.~6 in Ref.~\cite{Grajek:2008jb}).  This is
to be compared with our IB result of $\langle \sigma_A v
\rangle_{e^+e^-} \alt 2 \times 10^{-23} \textrm{cm}^3
\text{s}^{-1}$ $(4 \times 10^{-24} \textrm{cm}^3
\text{s}^{-1})$ at 100 GeV (1 TeV), again assuming an NFW profile.
(The results of Ref.~\cite{Grajek:2008jb} are very similar to those of
Ref.~\cite{Borriello:2008gy}, though weaker than those of
Ref.~\cite{Hooper:2008zg}, in which less conservative assumptions were
made.)
Note that these synchrotron studies assume annihilation to $W^+W^-$
(or $\bar{q} q$) which then decay to electrons, rather than direct
annihilation to $e^+e^-$.  Therefore, these electrons are not at the
DM mass, and have instead a broad distribution of energies centered on
the $W$ mass.  The synchrotron analyses
in~\cite{Grajek:2008jb,Borriello:2008gy,Hooper:2008zg} thus serve only
as an interesting reference point for our work, and not as a direct
comparison.

References~\cite{Cirelli:2008pk,ArkaniHamed:2008qn,Pospelov:2008jd,Nelson:2008hj,Cholis:2008qq,Bai:2008jt,Fox:2008kb}
have recently proposed models in which DM annihilates directly to
charged leptons, with cross sections well above that expected for a
thermal relic.
This may account for anomalies in cosmic ray spectra from PAMELA, HEAT and
ATIC, gamma-ray measurements from EGRET, and microwave signals from
the Wilkinson Microwave Anisotropy Probe, all of which seem to require more electrons and positrons than
can be explained otherwise.  Our bounds on $\langle \sigma_A v
\rangle_{l^+l^-}$ will directly constrain the allowed parameter space
for these types of DM models.

We expect the sensitivity of the IB bounds to be improved by
forthcoming data from the Fermi-GLAST
experiment~\cite{Morselli:2004ke,Baltz:2008wd,Jeltema:2008hf}.
Improved point source subtraction enabling the diffuse background to
be reduced, together with better energy and angular resolution and high
statistics measurements, will enable stronger limits to be placed on
all DM annihilation processes that produce gamma rays in the measured
energy range.

\section{Conclusions}

Dark matter annihilation into charged particles will necessarily be
accompanied by gamma rays.  Internal bremsstrahlung from final state
charged particles can produce hard gamma rays, close to the endpoint
defined by $E_\gamma =m_\chi$, with an approximately model-independent
spectrum.
Using galactic gamma-ray data, we have calculated upper limits on the
dark matter annihilation cross section to $e^+e^-$ and other charged
leptons.  We have made conservative assumptions about the
astrophysical inputs, and demonstrated how our derived bounds would be
strengthened if the galactic halo has a steeper density profile than
assumed.
The upper bound on the annihilation cross section into $e^+e^-$ is
weaker than that for the ideal $\gamma\gamma$ final state by only a
factor of $\alt 10^2$. For a wide range of masses, our upper bound on
$\langle \sigma_A v \rangle_{e^+e^-}$ is stronger than the bound on
the total cross section defined by neutrinos, the least detectable
final state.
Compared with recent constraints on DM annihilation cross sections
based upon synchrotron radiation, the internal bremsstrahlung
constraints on $\langle \sigma_A v \rangle_{e^+e^-}$ are broadly
comparable in strength.  However, synchroton emission depends strongly
on poorly known astrophysical inputs, such as galactic magnetic field
strengths.  In comparison, the normalization and spectrum of IB
radiation is fixed, independent of any astrophysical inputs, and is thus an 
extremely clean technique.


\medskip

{\bf Acknowledgments:}
We thank Gianfranco Bertone, Greg Mack, Stefano Profumo and Hasan
Yuksel for helpful discussions, and John Beacom for detailed comments
on the manuscript.
NFB was supported by a University of Melbourne Early Career
Research Grant, and TDJ by the Commonwealth of Australia.


\end{document}